\documentclass[groupedaddress, showpacs, superscriptaddress]{revtex4-2}
\usepackage{amssymb,amsmath,amsthm,color,graphicx,times,graphicx}
\usepackage{hyperref}
\usepackage{subfigure}
\usepackage{times}
\usepackage{bbold}
\usepackage{color}
\usepackage{array}
\usepackage{sidecap}
\usepackage[french]{babel}
\usepackage{array,multirow,makecell}
\usepackage[table]{xcolor}

\newcommand{\ket}[1]{|{#1}\rangle}

\usepackage{orcidlink}
\theoremstyle{plain}

\theoremstyle{definition}

\begin{document}
\title{Cyclic Quantum Teleportation of Two-Qubit Entangled States by using Six-Qubit Cluster State and Six-Qubit Entangled State}

\author{A. Slaoui \orcidlink{0000-0002-5284-3240}}\email{abdallah.slaoui@um5s.net.ma}\affiliation{LPHE-Modeling and Simulation, Faculty of Sciences, Mohammed V University in Rabat, Rabat, Morocco.}\affiliation{Centre of Physics and Mathematics, CPM, Faculty of Sciences, Mohammed V University in Rabat, Rabat, Morocco.}
\author{M. El Kirdi}\affiliation{LPHE-Modeling and Simulation, Faculty of Sciences, Mohammed V University in Rabat, Rabat, Morocco.}
\author{R. Ahl Laamara}\affiliation{LPHE-Modeling and Simulation, Faculty of Sciences, Mohammed V University in Rabat, Rabat, Morocco.}\affiliation{Centre of Physics and Mathematics, CPM, Faculty of Sciences, Mohammed V University in Rabat, Rabat, Morocco.}
\author{M. Alabdulhafith}
\affiliation{Department of Information Technology, College of Computer and Information Sciences, Princess Nourah bint Abdulrahman University, P.O. Box 84428, Riyadh 11671, Saudi Arabia.}
\author{S. A. Chelloug}\affiliation{Department of Information Technology, College of Computer and Information Sciences, Princess Nourah bint Abdulrahman University, P.O. Box 84428, Riyadh 11671, Saudi Arabia.}
\author{A. A. Abd El-Latif}\email{aabdellatif@psu.edu.sa}\affiliation{EIAS Data Science Lab, College of Computer and Information Sciences, and Center of Excellence in Quantum and Intelligent Computing, Prince Sultan University, Riyadh 11586, Saudi Arabia.}
\affiliation{Department of Mathematics and Computer Science, Faculty of Science, Menoufia University, Shebin El-Koom 32511, Egypt.}

\begin{abstract}
Cyclic quantum teleportation schemes requires at least the existence of three collaborators acting all as senders and receivers of quantum information, each one of them has an information to be transmitted to the next neighbour in a circular manner. Here, new cyclic quantum teleportation scheme is proposed for perfectly transmitting cyclically three arbitrary unknown two-qubit states ($\alpha$, $\beta$ and $\gamma$) among the three collaborators. In this scheme, Alice can send to Bob the quantum information contained in her two-qubit state $\alpha$ and receive from Charlie the quantum information contained in the two-qubit state in his possession $\gamma$ and similarly, Bob can transmit to Charlie the quantum information contained in his two-qubit state $\beta$ through a quantum channel of twelve-qubit state consisting of a six-qubit cluster state and a six-qubit entangled state by sequentially and cyclically performing Bell state measurements. Subsequently, each one of the three participants can afterwards retrieve his own desired two-qubit state using classical channel and by performing appropriate unitary Pauli operators  and we have shown that our proposed scheme performs efficiently.
\par
\vspace{0.25cm}
\par
\end{abstract}
\date{\today}

\maketitle
\section{ Introduction}
In the field of secure transmission of confidential information, the quantum teleportation (QT) is considered as protocol that allows the transfer of unknown quantum information, not only indecipherable but also imperceptible between two remote entities without loss of value. This opens the door to secure communication protocols over long distances and thus offers a reliable solution for the future quantum internet \cite{Kimble,Slaoui2023,Ikken2023}. The first scientific scheme that discusses QT as a protocol of secure transmission of quantum information contained in an unknown quantum state, was first developed by Bennet $et$ $ al$. \cite{Bennet}, where they found that quantum information from an unknown quantum state can be perfectly transmitted from a sender, say Alice, to a distant receiver, say Bob, via a maximally pre-shared entangled quantum channel, without physical transfer of the discussed system itself. In a more convincing way, after the measurement the unknown quantum state of the discussed physical system disappears from the transmitting entity to appear in the more distant receiving entity at the same time, without loss of information, and without taking into account the distance between the two entities using classical communication and quantum entanglement \cite{Brassard,Steane,Horodecki,Einstein}. Without this recent discovery, which has become in our time the most important principle of quantum mechanics, today we would not be talking about something called QT of an unknown quantum state according to the principles of quantum mechanics at the time, which considered something like this forbidden and impossible i.e. Heisenberg's principle of uncertainty. For this reason, QT is the incipient and direct result of this great discovery and is therefore the one who can better explain and provide an impressive explanation for the phenomenon of quantum entanglement  and the huge success that we can achieve if we are lucky enough to have it maximally among quantum entities \cite{Ma1,Gisin,Heo,Chen1,Vaidman,Ming}.\par

Experimentally, several physical systems have been transmitted with encouraging results
\cite{Riebe,Bouwmeester,Verma,Ma2,Li3,Na,Pirandola,Boschi}. Since QT is essentially a transmission protocol, it was obvious that most of attention will be focused on its effectiveness according to the different systems studied and the possibility of continuing its success in the presence of new constraints, for example, case of long distance \cite{Barrett} or also the quantum channel pre-shared between sender and receiver, such as Einstein-Podolsky-Rosen (EPR) state \cite{Rigolin}, Greenberger-Horner-Zeilinger (GHZ) state \cite{Dong, Espoukeh}, GHZ-like state \cite{Zhang, Nandi, Yuan}, Werner (W) state \cite{ Zuo, Cao}, W-like state \cite{Man} and cluster state \cite{Liu, Li, Li2, Sisodia}. Building on prior research \cite{Cheung2009}, this work investigates faithful teleportation using an arbitrary shared quantum state (qubit) between Alice and Bob. The authors introduce a general criterion to determine if this channel can teleport any $d$-qubit state accurately, regardless of the specific channel state. This new method offers a significant advantage; it's simpler to implement in experiments compared to existing protocols. A novel d-qudit teleportation protocol is presented in Ref.\cite{Zhang2007}. This scheme utilizes the tensor product state of $n$-generalized
	Bell states as the quantum channel, allowing for faithful state
	transmission. Additionally, the implementation of multi-qubit quantum
	teleportation and a deterministic secure communication protocol under
	the control of multiple agents are reported in \cite{Zhang2005}. Practically, parts responsible for QT can be extended to other parts with more tasks as needed as possible. The extension of QT to multipartite QT is known as controlled quantum teleportation (CQT) \cite{Karlsson}, which is an integral part of quantum teleportation networks \cite{Vanloock} and could be the  cornerstone of the future quantum internet. In 2013, an original novel protocol for bidirectional controlled quantum teleportation (BCQT) was firstly presented by Zha $et$ $al$. \cite{Zha}, where Alice and Bob decides together to teleport their quantum states to each other under the supervision of a controller, say Charlie using five-qubit cluster states. Interesting, Yuan and Zhang \cite{Yuan2023} have suggested an asymmetric BQT scheme feasible with present experimental technologies by using a four-qubit cluster state and a Bell state as a quantum channel. In addition, Wanbin et al.\cite{Zhang2020} have theoretically proposed three BQCT schemes, allowing bidirectional teleportation between two participants with another as the controller, by utilizing seven-qubit maximally entangled states. BCQT has been well received by researchers in the field of information transmission and has been applied using many different entangled pre-shared quantum channel, as for example GHZ states \cite{Hassanpour}, five-qubit entangled states \cite{Chen2}, five-qubit composite GHZ-Bell states \cite{Li4}, six-qubit entangled states \cite{Duan}, seven-qubit entangled states \cite{Duan2}, eight-qubit entangled state \cite{Zhang3} and multi-qubit entangled states \cite{Zadeh}. In 2017, an other novel protocol for cyclic quantum teleportation (CYQT) was originally proposed by Chen $et$ $al$. \cite{Chen4}, where three collaborators Alice, Bob and Charlie all decides to teleport their single-qubit quantum states among themselves using a six-qubit entangled state. Immediately after, Sang \cite{Sang} proposed a controlled CYQT protocol introducing a fourth co-operator playing the role of supervisor David, implying that Alice, Bob and Charlie could exactly share their quantum states if David gave his consent \cite{Ting,Wang,Li5,Chen}. Recently, Verma \cite{Verma1} suggested a new scheme for CYQT, where four co-operators Alice, Bob, Charlie and David can cyclically teleport their single-qubit state among themselves by using two G-states.\par

In our work, an efficient cyclic quantum teleportation of three arbitrary two-qubit states using a six-qubit cluster state and a six-qubit entangled state is proposed. In our scheme, Alice has an unknown two-qubit state containing two qubits $\alpha_0$ and $\alpha_1$, Bob has a second unknown two-qubit state containing qubits $\beta_0$ and $\beta_1$, and the third collaborator Charlie has also a third unknown two-qubit state containing two qubits $\gamma_0$ and $\gamma_1$. The goal of our CYQT scheme is to teleport the two-qubit state in Alice's possession to his near neighbour Bob, and Bob must transmit his two-qubit state to his neighbour Charlie and also the third co-operator Charlie can teleport his unknown two-qubit state to Alice. To achieve this, the three parties must first share a quantum channel containing twelve-qubit distributed among the three collaborators, each of them has four qubits. Next, each collaborator must to perform Bell state measurements on his two qubits to be transmitted and his two out of these four qubits of the quantum channel. Finally, each one of them might to perform  suitable Pauli unitary operators on his resulting state to obtain the original state and this after, of course, receiving measurement results by classical channel from the responsible sender.\par

The outline of this article is organized as follows: In Section 2, we introduced the cyclic quantum teleportation scheme with a mention of the construction of the used pre-shared entangled twelve-qubit quantum channel, the three target states to be teleport for Alice to Bob, from Bob to Charlie and from Charlie to Alice respectively. Section 3 details the execution of cyclic quantum teleportation scheme of the three arbitrary two-qubit entangle entangled state. Section 4 analysis the amout data of classical information, the success probability and the protocol performance by calculating the efficiency factor of the proposed CYQT scheme. Concluding remarks are given in Section 5.

\section{Three arbitrary two-qubit states to be transmitted and selected twelve-qubit entangled quantum channel}

In this section, we aim to represent the quantum channel used, the unknown selected target states and the CYQT protocol steps. Suppose that there are three participants, employing twelve-qubit entangled states combining a six-qubit cluster state and a six-qubit entangled state as a quantum channel. Alice has an unknown two-qubit state $\ket{\psi}_{\alpha}$ to teleport to Bob, Bob has a second unknown two-qubit state $\ket{\psi}_{\beta}$ to teleport to Charlie, and similarly Charlie has a third unknown two-qubit state $\ket{\psi}_{\gamma}$ to teleport to Alice through a quantum channel. The quantum channel linking the three participants Alice, Bob and Charlie take the form
\begin{equation}
	\ket{\psi}_{123456789101112} = \ket{C_{6}}_{123456} \otimes \ket{\psi_{6}}_{789101112},
\end{equation}
the six-qubit cluster state $\ket{C_{6}}$ used in the quantum channel above has the form
\begin{equation}
	\ket{C_{6}} = \dfrac{1}{2}\{ \ket{000 000} + \ket{000 111} + \ket{111 000} +\ket{111 111}\},
\end{equation}
and the six-qubit entangled state $\ket{\psi_{6}}$ can be described as follows
\begin{align}
	\ket{\psi_{6}} = \dfrac{1}{4}\{ \ket{000 000}& + \ket{000 001}  +...+ \ket{000 110} + \ket{000 111} \nonumber\\
	& +\ket{111 000} +\ket{111 001} +...+ \ket{111 110} +\ket{111 111}\}.
\end{align}

\subsection{Construction of the quantum channel}
\begin{figure}
\begin{center}
	\includegraphics[scale=0.25]{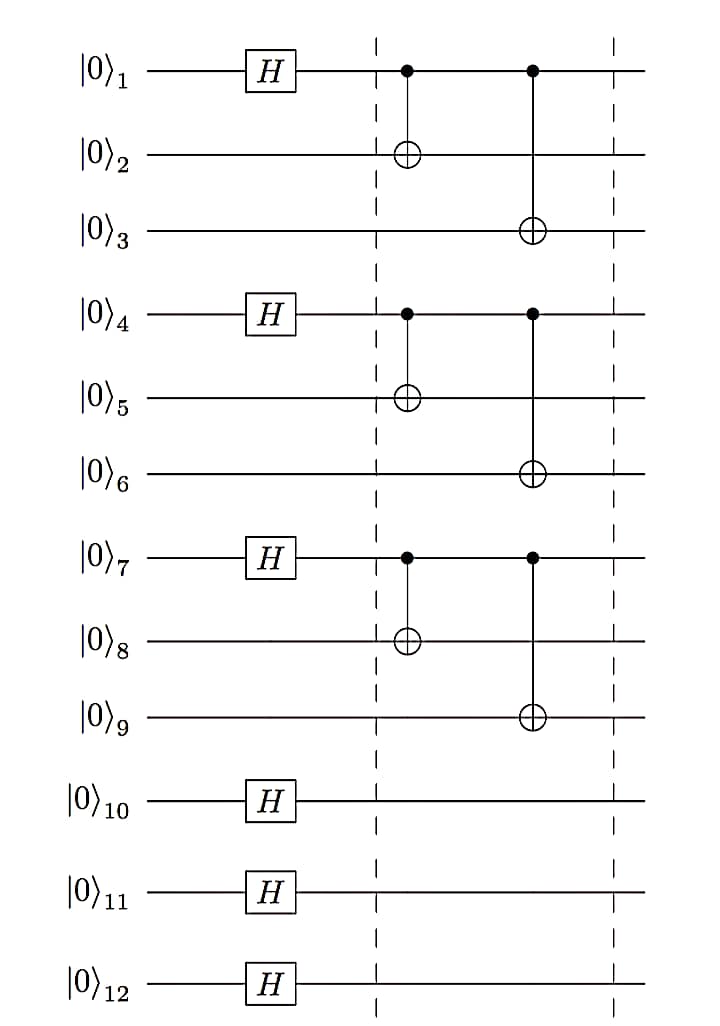}
	\caption{Quantum circuit illustrating the construction of the twelve-qubit quantum channel combining a six-qubit cluster state and a six-qubit entangled state using H-gate and CNOT-gate.}\label{Fig1}
\end{center}
\end{figure}
The quantum channel in the simplest case is a twelve-qubit entangled state shared between the three participants Alice, Bob and Charlie, which can be realized by applying Hadamard (H) and Controlled-Not (CNOT) quantum gates (see the Fig.(\ref{Fig1})). Firstly, the product state of twelve-qubit state initialized to $\ket{0}$ is used as the input state $\ket{\phi_{0}}$ in the quantum circuit, which can be expressed as
\begin{equation}
	\ket{\phi_{0}} = \ket{0}_{1} \otimes \ket{0}_{2}  \otimes ... \otimes \ket{0}_{12} ,
\end{equation}	
then, performing H-gates on qubits (1, 4, 7, 10, 11, 12), the input state $\ket{\phi_{0}}$
is converted to $\ket{\phi_{1}}$ as follows	
\begin{align}
	\ket{\phi_{1}}& = \dfrac{1}{8}\{\ket{000000} + \ket{000100} + \ket{100000}+ \ket{100100}\}_{123456}\otimes \{ \ket{000} + \ket{100}\}_{789}  \nonumber\\
	& \otimes \{\ket{000} + \ket{001} + \ket{010} + \ket{011} + \ket{100} + \ket{101} + \ket{110} + \ket{111}\}_{101112},
\end{align}
after that, several CNOT-gates need to be performed on the following qubit pairs (1;2), (1;3), (4;5), (4;6), (7;8) and (7;9), qubits 1, 4, 7 control qubits (2;3), (5;6) and (8;9) respectively. Afterwards, the quantum channel takes the form (1). Where Alice holds qubits (5,6,7,10), Bob holds qubits (1,8,9,11) and Charlie holds qubits (2,3,4,12). The twelve-qubit entangled state can be rewritten as
\begin{align}
	\ket{\psi_{12}}& = \dfrac{1}{8}\{\ket{000000} + \ket{000111} + \ket{111000}+ \ket{111111}\}_{B_4C_1C_3C_4A_1A_3}\otimes \{ \ket{000} + \ket{111}\}_{A_4B_1B_3}  \nonumber\\
	& \otimes \{\ket{000} + \ket{001} + \ket{010} + \ket{011} + \ket{100} + \ket{101} + \ket{110} + \ket{111}\}_{A_2B_2C_2},
\end{align}
\begin{figure}
	
\end{figure}
\subsection{Description of the proposed CYQT scheme }
\begin{figure}
	\begin{center}
		\includegraphics[scale=0.6]{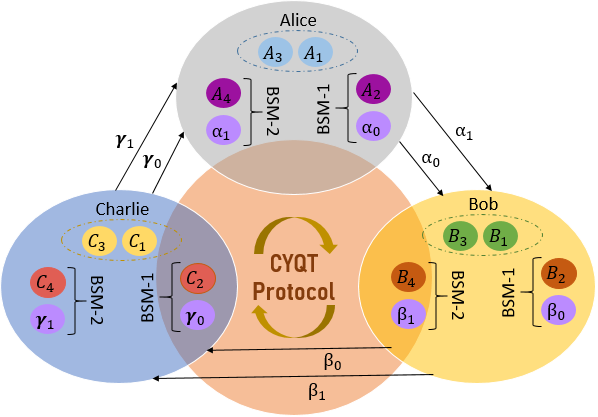}
	\caption{Description of the protocol of CYQT, each one of the three observers Alice, Bob and Charlie has a two-qubit state ($\alpha_0$, $\alpha_1$), ($\beta_0$,$\beta_1$) and  ($\gamma_0$,$\gamma_1$) respectively to teleport among themselves via quantum channel and by performing BSM.}\label{Fig2}
	\end{center}
\end{figure}

In this scheme, each one of three participants has an arbitrary unknown two-qubit state to teleport to the next neighbour. Alice has an unknown two-qubit state $\ket{\psi}_{\alpha}= a_0 \ket{00}_{\alpha_0\alpha_1} + a_1 \ket{11}_{\alpha_0\alpha_1}$ to teleport to Bob, Bob has a second unknown two-qubit state $\ket{\psi}_{\beta}= b_0 \ket{00}_{\beta_0\alpha_1} + b_1 \ket{11}_{\beta_0\beta_1}$ to teleport to Charlie, while Charlie has a third unknown two-qubit state $\ket{\psi}_{\gamma}= c_0 \ket{00}_{\gamma_0\gamma_1} + c_1 \ket{11}_{\gamma_0\gamma_1}$ to teleport to Alice through the quantum channel discussed above. Where $a_0$, $a_1$, $b_0$, $b_1$, $c_0$, $c_1$ are arbitrary complex coefficients with $|a_0|^2+|a_1|^2=1$, $|b_0|^2+|b_1|^2=1$ and $|c_0|^2+|c_1|^2=1$. The description of our CYQT scheme is shown in Figure \ref{Fig2}. 
Alice wants to transmit $\alpha_0$ and $\alpha_1$ to Bob, Bob wants to transmit his own two qubits $\beta_0$ and $\beta_1$ to Charlie, and similarly Charlie wants to transfer the two qubits in its possession $\gamma_0$ and $\gamma_1$ to Alice.
\subsection{CYQT protocol steps}
The three participants Alice, Bob and Charlie, are spatially separated and entangled by a quantum channel, which means that the quantum information will be transmitted using the pre-shared quantum channel as well as other classical communication channels. The steps used to make our CYQT scheme are organized as follows:
\begin{itemize}
	\item After combining the three arbitrary two-qubit states to be transmitted with the twelve-qubit quantum channel, Alice carries out Bell state measurements (BSM) on qubits $\alpha_0$ and $A_2$, and will send to Bob the result of her measurement by classical communication channel.
	\item Bob carries out BSM on qubits $\beta_0$ and $B_2$, and will send his BSM results to Charlie by classical communication channel.
	\item Charlie carries out BSM on qubits $\gamma_0$ and $C_2$, and will send to Alice the result of his measurement by classical communication channel.
	\item Alice carries out a second BSM on qubits $\alpha_1$ and $A_4$, and will send to Bob the result of his last measurement by classical communication channel.
	\item Charlie carries out his second BSM on qubits $\gamma_1$ and $C_1$, and will send to Alice the result of his measurement by classical communication channel.
	\item Bob ends the cycle of Bell state measurements, by carrying out BSM on qubits $\beta_1$ and $B_4$, and will send his BSM results to Charlie by classical communication channel.
\end{itemize}
Based on the received results, Alice, Bob and Charlie can recover the initial two-qubit quantum state send by Charlie, Alice and Bob respectively by applying appropriate unitary transformations ($I$, $X$, $i$$Y$,$Z$).

\section{Cyclic quantum teleportation of three arbitrary two-qubit states among the three users}

\begin{figure}
	\begin{center}
		\includegraphics[scale=0.4]{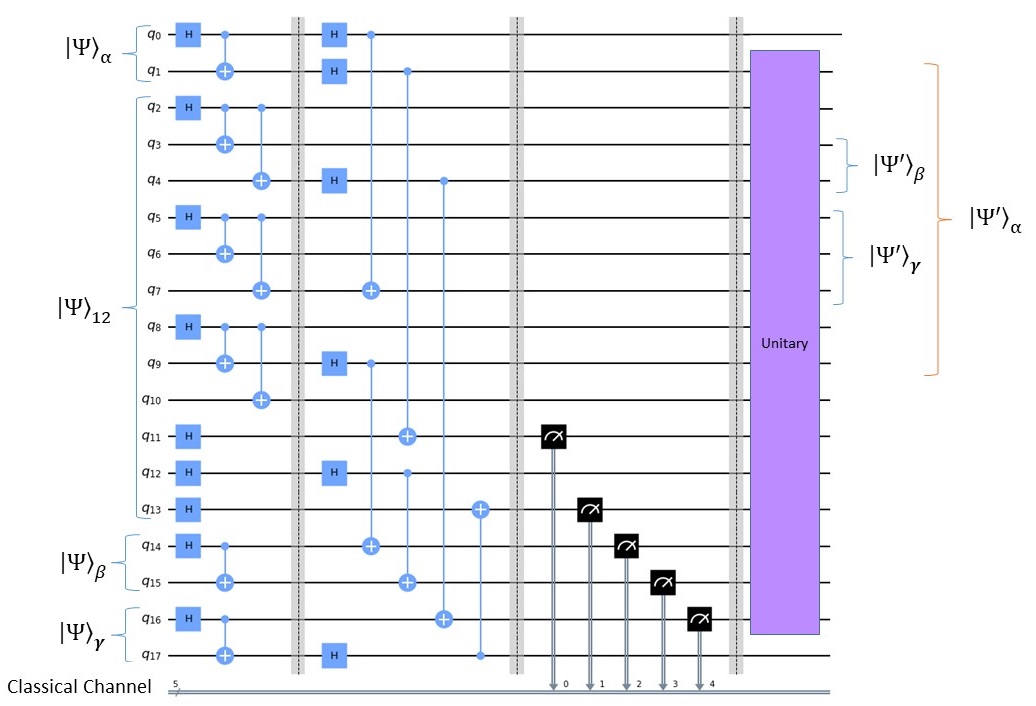}
	\caption{Quantum circuit illustrating the scheme of the proposed CYQT protocol of three two-qubit state $\ket{\psi}_{\alpha}$, $\ket{\psi}_{\beta}$ and $\ket{\psi}_{\gamma}$ using a twelve-qubit state $\ket{\psi}_{12}$ combining a six-qubit cluster state and a six-Bell state. The double lines decodes the classical wires and $"U"$ decode Pauli unitary operations I, X, $i$Y and Z. The output states $\ket{\psi^{'}}_{\alpha}$, $\ket{\psi^{'}}_{\beta}$ and $\ket{\psi^{'}}_{\gamma}$ represented the recovered states at the end of the teleportation protocol.}\label{Fig3}
	\end{center}
\end{figure}

To teleport the three arbitrary two-qubit state among the three collaborators Alice, Bob and Charlie as described in Figure \ref{Fig2}, one has to follow the CYQT protocol steps well detailed in subsection II.3 (see the Fig.(\ref{Fig3})). The combined state of the whole subsystems can be expressed as
\begin{align}
	\left|\psi\right>_{\alpha_0\alpha_1\beta_0\beta_1\gamma_0\gamma_1B_4C_1C_3C_4A_1A_3A_4B_1B_3A_2B_2C_2} = \left|\psi \right>_{\alpha}\otimes\left|\psi\right>_{\beta} \otimes \left|\psi\right>_{\gamma} \otimes \left|\psi_{12}\right>,
\end{align}	
\textbf{Step 1:} To implement the cyclic quantum teleportation protocol, Alice takes the first step by applying a BSM on her qubits $\alpha_0$ and $A_2$, the BSM basis is given by,
\begin{align}
	\left|\Phi^{\pm}\right> = \dfrac{1}{\sqrt{2}}\{\left|00\right> \pm \left|11\right>\}, \hspace{1cm} \left|\Psi^{\pm}\right> = \dfrac{1}{\sqrt{2}}\{\left|01\right> \pm \left|10\right>\},
\end{align}	
After measurement, if Alice's BSM result was $\ket{\Phi^{+}}_{\alpha_0A_2}$, the 
state of the remaining sixteen subsystems suddenly collapsed to
\begin{align}
	&\dfrac{\sqrt{2}}{4}\{a_0\ket{0}_{\alpha_1}+a_1\ket{1}_{\alpha_1}\}\otimes \left|\psi\right>_{\beta}\otimes \left|\psi\right>_{\gamma}\otimes \ket{C_6}_{B_4C_1C_3C_4A_1A_3} \nonumber\\
	&\otimes \{\ket{000}+\ket{111}\}_{A_4B_1B_3}\otimes \{\ket{00}+\ket{01}+\ket{10}+\ket{11}\}_{B_2C2}
\end{align}
\textbf{Step 2:} Immediately, Bob realize the second step in this protocol by carrying out a BSM on his qubits $\beta_0$ and $B_2$ in the same basis suggested in $Eq.(8)$. Suppose that Bob's BSM result was also $\left|\Phi^+\right>_{\beta_0B_2}$, the state of the remaining fourteen subsystems suddenly collapsed to
\begin{align}
	&\dfrac{1}{2}\{_0\ket{0}_{\alpha_1}+a_1\ket{1}_{\alpha_1}\}\otimes\{b_0\ket{0}_{\beta_1}+b_1\ket{1}_{\beta_1}\}\otimes \left|\psi\right>_{\gamma}\otimes\nonumber\\ &\ket{C_6}_{B_4C_1C_3C_4A_1A_3} 
	\otimes \{\ket{000}+\ket{111}\}_{A_4B_1B_3}\otimes \{\ket{0}+\ket{1}\}_{C2}
\end{align}
\textbf{Step 3:} Basing on the same BSM basis proposed in $Eq.(8)$, the third party Charlie ends the first cycle of the discussed CYQT by performing a BSM on his qubits $\gamma_0$ and $C_2$. If Charlie's BSM result was $\left|\Phi^+\right>_{\gamma_0C_2}$, the state of the remaining twelve subsystems collapsed to
\begin{align}
	\dfrac{\sqrt{2}}{2}\{a_0\ket{0}_{\alpha_1}+&a_1\ket{1}_{\alpha_1}\}\otimes\{b_0\ket{0}_{\beta_1}+b_1\ket{1}_{\beta_1}\}\otimes \{c_0\ket{0}_{\gamma_1}+c_1\ket{1}_{\gamma_1}\}\otimes\nonumber\\ &\ket{C_6}_{B_4C_1C_3C_4A_1A_3} 
	\otimes \{\ket{000}+\ket{111}\}_{A_4B_1B_3}
\end{align}
\textbf{Step 4:} In the following steps, the three parties will realize the second cycle of the discussed CYQT. In this step, Alice carry out a second BSM on her qubits $\alpha_1$ and $A_4$. If Alice's new BSM result was $\left|\Phi^+\right>_{\alpha_1A_4}$, the state of the remaining ten subsystems would collapse into
\begin{align}
	\{b_0\ket{0}_{\beta_1}+b_1\ket{1}_{\beta_1}\}\otimes \{c_0\ket{0}_{\gamma_1}+c_1\ket{1}_{\gamma_1}\}\otimes \ket{C_6}_{B_4C_1C_3C_4A_1A_3} 
	\otimes \{a_0\ket{00}+a_1\ket{11}\}_{B_1B_3}
\end{align}
at the end of this step, it is clear that Bob received Alice's two-qubit state.\\

\hspace{-0.55cm}\textbf{Step 5:} In this sense, Charlie re-perform a second BSM on his qubits $\gamma_1$ and $C_4$. If Charlie's new BSM result was $\left|\Phi^+\right>_{\gamma_1C_4}$, the state of the remaining height qubits collapsed to
\begin{align}
	\dfrac{\sqrt{2}}{2}\{b_0\ket{0}_{\beta_1}+b_1\ket{1}_{\beta_1}\}\otimes &\{c_0\ket{00000} +c_1 \ket{00011} + c_0\ket{11100}+ c_1\ket{11111}\}_{B_4C_1C_3A_1A_3}\nonumber\\
	&\otimes \{a_0\ket{00}+a_1\ket{11}\}_{B_1B_3}
\end{align}
\textbf{Step 6:} Finally, Bob ends the second cycle of our proposed CYQT by performing the last BSM on his qubit $\beta_1$ and $B_4$. If Bob's last BSM result was $\ket{\Phi^+}_{\beta_1B_4}$, the remaining six qubits might collapse to the state below
\begin{align}
	\{b_0c_0\ket{0000} +b_0c_1 \ket{0011} + b_1c_0\ket{1100}+ b_1c_1\ket{1111}\}_{C_1C_3A_1A_3}
	\otimes \{a_0\ket{00}+a_1\ket{11}\}_{B_1B_3}.
\end{align}
This final state can be expressed as
\begin{align}
	\ket{\psi}_{C_1C_3A_1A_3B_1B_3}= \{\ket{\psi^{'}}_{\beta}\}_{C_1C_3}\otimes\{\ket{\psi^{'}}_{\gamma}\}_{A_1A_3}\otimes\{\ket{\psi^{'}}_{\alpha}\}_{B_1B_3},
\end{align} 
with
\begin{align}
	&\ket{\psi^{'}}_{\beta}= \{b_0\ket{00} +b_1 \ket{11}\}\nonumber\\
	&\ket{\psi^{'}}_{\gamma} = \{c_0\ket{00} +c_1 \ket{11}\}\nonumber\\
	&\ket{\psi^{'}}_{\alpha}=  \{a_0\ket{00}+a_1\ket{11}\},
\end{align}
which are exactly the desired three two-qubit arbitrary states sends respectively by Bob to Charlie, by Charlie to Alice and by Alice to Bob. This means that receivers will not need to do any conversion to get the original sent states. This in the case when the BSMs results were respectively $\ket{\Phi^+}_{\alpha_0A_2}$, $\ket{\Phi^+}_{\beta_0B_2}$, $\ket{\Phi^+}_{\gamma_0C_2}$, $\ket{\Phi^+}_{\alpha_1A_4}$, $\ket{\Phi^+}_{\gamma_1C_4}$ and $\ket{\Phi^+}_{\beta_1B_4}$. For other possible BSM outcomes, receivers will have to apply appropriate transformations before they can exactly retrieve the original ones. Due to classical communication, receivers will be able to know exactly which Pauli unitary operators will be necessary to recover the exactly ones. To be precise, there would be $2^{12}=4096$ possible measurement states (see the Table (\ref{Tabl1})). Each one of them will have a certain series of appropriate unitary transformations that must be applied before the original transmitted information states can be reliably recuperated. We will not go over all of them, first because they are many, and secondly, because necessary transformations will be clear and intuitive. But despite that, and for more clarification, we discussed the first four of them. In table below, we can find the first four possible BSM result cases and their necessary conversions. Whatever the result of applied measurements, Alice, Bob and Charlie will be able to recover the original transmitted information states and thus we can be sure that the proposed CYQT protocol runs perfectly in all cases.

\begin{table}[h!]
	\centering
	\renewcommand{\arraystretch}{1.3}
	\begin{tabular}{ p{1.3cm} p{1.3cm} p{1.3cm} p{1.3cm} p{1.5cm} p{1.5cm}| p{1.6cm} p{1.6cm} p{1.6cm} }
		\hline
		\multicolumn{6}{c|}{\textbf{Bell State Measurements (BSM)} } &		\multicolumn{3}{c}{\textbf{Pauli Unitary Operators }}\\
		\hline
		Alice&Bob&Charlie&Alice&Charlie&Bob&Alice&Bob&Charlie\\
		\hline
		\multirow{4}{4em}{$\ket{\Phi^+}_{\alpha_0A_2}$} &		\multirow{4}{4em}{$\ket{\Phi^+}_{\beta_0B_2}$}&		\multirow{4}{4em}{$\ket{\Phi^+}_{\gamma_0C_2}$}&		\multirow{4}{4em}{$\ket{\Phi^+}_{\alpha_1A_4}$}&		\multirow{4}{4em}{$\ket{\Phi^+}_{\gamma_1C_4}$} & $\ket{\Phi^+}_{\beta_1B_4}$&$I_{A_1}\otimes I_{A_3}$ & $I_{B_1}\otimes I_{B_3}$ & $I_{C_1}\otimes I_{C_3}$\\
		&&&&	 & $\ket{\Phi^-}_{\beta_1B_4}$&$I_{A_1}\otimes I_{A_3}$ & $I_{B_1}\otimes I_{B_3}$ & $Z_{C_1}\otimes I_{C_3}$\\
		&&&&	 & $\ket{\Psi^+}_{\beta_1B_4}$&$I_{A_1}\otimes I_{A_3}$ & $I_{B_1}\otimes I_{B_3}$ & $X_{C_1}\otimes X_{C_3}$\\
		&&&&   	 & $\ket{\Psi^-}_{\beta_1B_4}$&$I_{A_1}\otimes I_{A_3}$ & $I_{B_1}\otimes I_{B_3}$ & $iY_{C_1}\otimes X_{C_3}$\\
		\hline
	\end{tabular}
\caption{First four results among "$4096$" possible BSM outcomes and Pauli unitary operators need to be applied by the three receivers to retrieve original states. $i$ is an imaginary unit, $X$, $Y$ and $Z$ are Pauli unitary operators and $I$ is the identity unitary operator.}
	\label{Tabl1}
\end{table}
\begin{figure}[hbtp] 
	{{\begin{minipage}[b]{1\linewidth}
				\centering
				\includegraphics[scale=0.5]{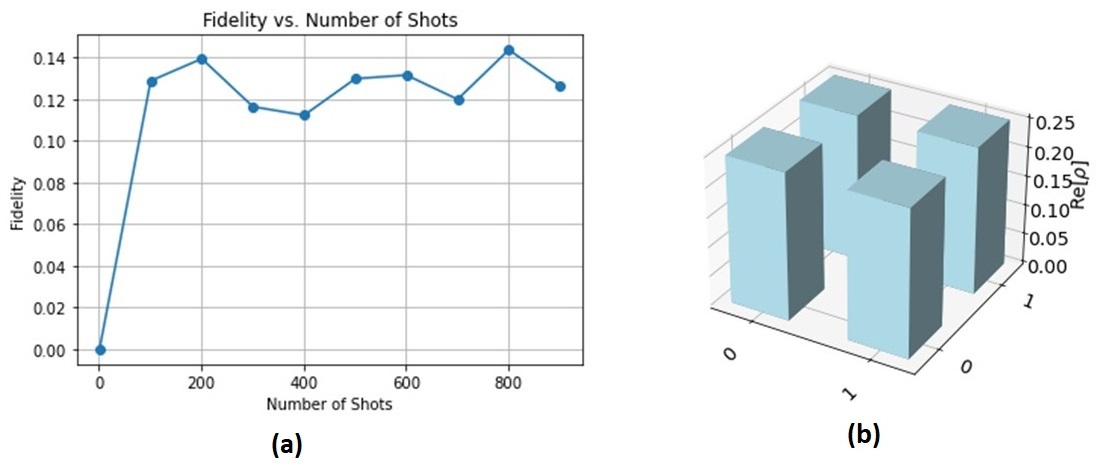}
	\end{minipage}}}
	\caption{Panel ($a$) presents the quantum circuit fidelity, giving information about the accuracy of quantum teleportation protocol. Panel ($b$) shows the Teleportation Density, showing the distribution and characteristics of quantum states after teleportation.}\label{Fig4}
\end{figure}

Using Qiskit, the code (\ref{Fig3}) prepares the quantum and classical registers. After defining a Toffoli gate matrix, it builds a quantum circuit that carries out several quantum operations. These measures correspond to the steps described above of the CYQT protocol. To transmit the quantum states, Alice, Bob, and Charlie each execute Bell State Measurements (BSM) on their individual qubits in the CYQT protocol. These procedures, which involve the use of CNOT gates, Hadamard gates, and measurements, appear in the circuit. A unitary gate that reflects the outcomes of the BSMs executed by the three parties is applied at the circuit's conclusion. In order to represent the successful teleportation of the initial states between the parties, the circuit collapses the quantum state into a final state. the circuit show how the quantum operations transfer to the steps in the article, where each step is a particular party executing a BSM and collapsing the quantum state correspondingly. As described in the attached paper, the circuit functions as an implementation of a quantum circuit for CYQT, allowing the teleportation of arbitrary quantum states between multiple parties.\par

The fidelity of the quantum circuit is shown along the $X$-axis from $0$ to $900$ in the graph \ref{Fig4}(panel ($a$)). The number of shots varies. Data points are common, appearing approximately every $100$ frames. On the $Y$-axis, the fidelity is shown as a range between $0.02$ and a little above $0.12$. The fidelity shows significant variation across different shot counts, with no clear increasing or negative tendency. Initially, fidelity increases significantly from zero shots to approximately one hundred. After about $300$ shots, it stabilizes in the upper half of the graph after fluctuating for a while. The fidelity reaches its maximum at around $700$ shots, or $0.12$. The fidelity, on the other hand, reaches its minimum approximately 0.04, 200 shots after the beginning increase. When you get over $300$ shots, fidelity seems to settle off even more. In the last $600–900$ shots, there is less variation, but there are still obvious fluctuations. It also means that the performance of the circuit can vary depending on a number of factors or quantum processes like decoherence. In general, there is not an obvious increase or decrease trend, even when fidelity stabilizes at the end, indicating to the detailed dynamics of the quantum system.\par

The left side of graph \ref{Fig4}(panel ($b$)) shows a three-dimensional grid that represents the density matrix components along the $z$-axis. This grid represents Alice's (the first sender) unknown state. Alice unkown variable was given a randowm value to represent the density matrix, we could repeat the process for Bob and Charlie, but with different values. Every block in the grid represents a matrix element, and the element's value is indicated by its height. All elements are non-negative and range from $0$ to $0.25$ on the $z$-axis scale, with a maximum value of $0.25$. The color blue, doesn't change continuously.

\section{Protocol success probability and efficiency}
As we have already shown, Alice, Bob and Charlie each have to apply two Bell state measurements, i.e. there will be a total of six Bell state measurements. So, we'll have 4096 possible Bell state measurements results. The probability of each result is given by 
\begin{equation}
	\hspace{2cm}	P_i = \left( \dfrac{1}{2^6}\right)^2 = \dfrac{1}{4096};\hspace{1cm} \{i=0, 1,...,4095\}
\end{equation}
$\bullet$ The total success probability can be expressed as
\begin{equation}
	P_{total} = \sum_{i=0}^{4095} P_i = 100 \%
\end{equation}
$\bullet$ The average lack of information $S$ can be expressed as
\begin{equation}
	S = - \sum_{i=0}^{4095}P_i\hspace{0.25cm}log\hspace{0.25cm}P_i = 12
\end{equation}
it means that results have all equal probabilities. To study the performance of communication protocols \cite{Kao}, we often calculate the efficiency factor defined by
\begin{equation}
	e\left(\%\right) = \dfrac{q_t}{q_s}
\end{equation}
where $q_t$ denotes the number of qubits have been teleportated during the execution of the protocol (here it was three arbitrary two-qubit state i.e. 6 teleported qubits). And, $q_s$ denotes the total number of qubits of the quantum channel used for teleportation (here we used a combination of a six-qubit cluster state and a six-qubit entangled state i.e. 12 entangled qubits), which means that the proposed protocol has an efficiency that can reach $50\%$.
\section{Concluding remarks}
In summary, an original and perfect protocol of cyclic quantum teleportation by using a combination of six-qubit cluster state and six-qubit entangled state as quantum channel is presented for the first time. Where, the three distant parties Alice, Bob and Charlie are not only senders of information but also receivers of information. Alice can send an arbitrary two-qubit state to Bob and receive an arbitrary two-qubit state from Charlie and Bob can transmit an arbitrary two-qubit state to Charlie by performing Bell state measurements and appropriate unitary Pauli transformations after receiving the Bell state measurement results sent by others by classical way. In our scheme, the proposed quantum channel can be constructed by applying Controlled-Not gate and Hadamard gate on a twelve-qubit states all initialized to $\left|0\right>$ which has been implemented experimentally. During the execution of the protocol, the three collaborators choose the same basis of Bell states to perform the necessary measurements on the qubits they have. In fact, each of the three parties must perform two Bell state measurements on their qubits, which means that there will be a total of six Bell state measurements, so there will be 4096 possible output states. All these cases have equal probabilities and the protocol offers a success probability of up to 100\%. Particular attention was paid to the performance of the proposed protocol by calculating the information transfer efficiency factor and it was observed that this efficiency can reach 50\%, which is a very important and honourable percentage. Finally, it is hoped that the results obtained and the proposed protocol will bring an additional advantage in the world of quantum communication and data transmission. This proposed protocol offers a novel and secure method for transmitting quantum information between three distant parties. It enables not only sending but also receiving information, thereby enhancing the efficiency of quantum communication. However, generalizing this approach to other quantum teleportation tasks (like CBCQT) or applying it to larger systems might pose challenges. Further investigation and analysis are necessary to determine the protocol's effectiveness in these scenarios. We will address this question in a future work.

{\bf Data availability:} The datasets used and/or analyzed during the current study available from the corresponding author on request.\par

{\bf Funding statement:} Princess Nourah bint Abdulrahman University Researchers Supporting Project number (PNURSP2024R407), Princess Nourah bint Abdulrahman University, Riyadh, Saudi Arabia.\par

{\bf Acknowledgements:} Princess Nourah bint Abdulrahman University Researchers Supporting Project number (PNURSP2024R407), Princess Nourah bint Abdulrahman University, Riyadh, Saudi Arabia. Also, Ahmed A. Abd El-Latif would like to thank Prince Sultan University for their support.

\end{document}